\documentclass[12pt,fullpage]{article}
\usepackage[english]{babel}
\usepackage{epsfig,epsf}
\usepackage{graphicx}
\usepackage{hyphenat}
\newcommand{\as}{{\alpha}_s}
\author{V.A. Abramovsky, N.V. Prikhod'ko \\
Novgorod State University, B. S.-Peterburgskaya Street 41, \\
Novgorod the Great, Russia, 117259}
\title{Cronin momentum behavior in saturation model for $p+A$, $d+A$, $A+A$ collisions}

\begin{document}
\maketitle

\abstract{In this paper we consider Cronin momentum behaviour for $p+A$, $d+A$ and $A+A$
collisions in saturation model. Our analysis shows that Cronin momentum behavior at different rapidities and energies,
can be related with scaling law using simple dimensional consideration.
Using exact numerical solution of Balitsky-Kovchegov
equation we show that although this dependence is slightly different for McLerran-Venugopalan and
Balitsky-Kovchegov definition of gluon distribution function in simple model in this case dependencies
is almost the same (i.e ratio of Cronin momentum calculated using these gluon distribution functions is big constant).
This can be used to experimentally distinguish this two variant of gluon distribution function definition in saturation model
and choose the right one.
}

\section{Introduction}
In our previous paper \cite{Abramovsky:2003xx} it was found that in saturation model Cronin momentum (i.e. momentum at which
Cronin ratio \cite{Cronin:zm} have maximum) has simple behavior in central
rapidity region which holds in $p+A$ and $A+A$ collisions
\begin{equation}
	\ln(q_C)=a+b\ln(\sqrt{s})
	\label{CM_simple}
\end{equation}
where parameters $a$ and $b$ are {\it approximately} defined as:
\begin{eqnarray}
\label{ab_abs}
	a=\frac{1}{3(2+\lambda)}\ln(A)+\ln(q^0_C)\,, \\\nonumber
	b=\frac{\lambda}{2+\lambda}=0.1304\,.
\end{eqnarray}
with geometric scaling parameter $\lambda=0.3$.
It was also showed that for different definitions(\ref{phi_cil_BK},\ref{phi_cil_MV}) of unintegrated gluon
distribution function slope $b$ is slightly different.

Relation (\ref{CM_simple}) was inspired by simple expression (\ref{qsat_simple}) for saturations scale $Q_s(x)$
based on geometric scaling effect and dimensional considerations (\ref{dim_cons}).
\begin{equation}
\log Q_s^2(Y)=\lambda Y + log(Q_s^2(Y_0)x_0^{\lambda})
\label{qsat_simple}
\end{equation}
\begin{equation}
Q_s^2\left(\frac{q_C}{\sqrt{s}}\right)=\beta q_C
\label{dim_cons}
\end{equation}
Since our calculation was done using certain model for dipole forward scattering amplitude and approximate expression for
cross section proposed in \cite{gribov} then it is most probably that (\ref{CM_simple}) does not holds if dipole forward scattering amplitude defined with
solid theoretical background will be taken into account. It could be shown below that this indeed true and simple formula
(\ref{CM_simple}) will be asymptotically restored only in high energy region.

There is mostly the only one way to define dipole forward scattering amplitude: Balitsky-Kovchegov evolution equation \cite{BK_equation},
which describes the evolution of the dipole forward scattering amplitude $N({\bf r},y)$ of a QCD dipole of transverse size
$|{\bf r}|$ with rapidity $Y=ln(1/x)$,
\begin{equation}
{dN(|{\bf r}|,Y)\over dy} = {(\as\, N_c)\over 2\pi^2}
\int d^2{\bf z}
{{\bf r}^2\over ({\bf r}-{\bf z})^2\, {\bf z}^2}
(N(|{\bf r}-{\bf z}|)+N(|{\bf z}|)- N(|{\bf r}|)-
N(|{\bf r}-{\bf z}|)N(|{\bf z}|))
\label{BK_eqn}
\end{equation}
here for simplicity we suppose that nucleus is cylindrical (i.e. we completely drop impact parameter ${\bf b}$
dependence from forward dipole scattering amplitude) and strong coupling constant $\as$ is fixed.
There is also many its derivatives but we'll leave them for now since most of them aren't established very well and bring to additional complexity in numerical calculation.
It was shown in series of papers \cite{BK_FKPP} that equation (\ref{BK_eqn}) belongs to Fisher-Kolmogorov-Petrovsky-Piscounov class \cite{FKPP} if $N(r,x)$
transformed to momentum space. It can be easily shown \cite{p_qsat_real} that instead of simple scaling law
(\ref{qsat_simple}) we will have the following one:
\begin{equation}
\log Q_s^2(Y)
=\bar\alpha\frac{\chi(\gamma_c)}{\gamma_c} Y-\frac{3}{2\gamma_c}\log Y\\
-\frac{3}{\gamma_c^2}
\sqrt{\frac{2\pi}{\bar\alpha\chi^{\prime\prime}(\gamma_c)}}\frac{1}{\sqrt{Y}}
+{\cal O}(1/Y)
\label{qsat_real}
\end{equation}
where $\chi(\gamma)=2\psi(1)-\psi(\gamma)-\psi(1-\gamma)$ is well known BFKL kernel, and $\gamma_c=0.6275...$ is the solution of equation $\chi({\gamma}_c)={\gamma}_c\chi({\gamma}_c)$.
It is oblivious that equation (\ref{CM_simple}) does not holds in this case but probably will be restored at high energy.

\begin{figure}
\resizebox{1\hsize}{!}{\epsffile{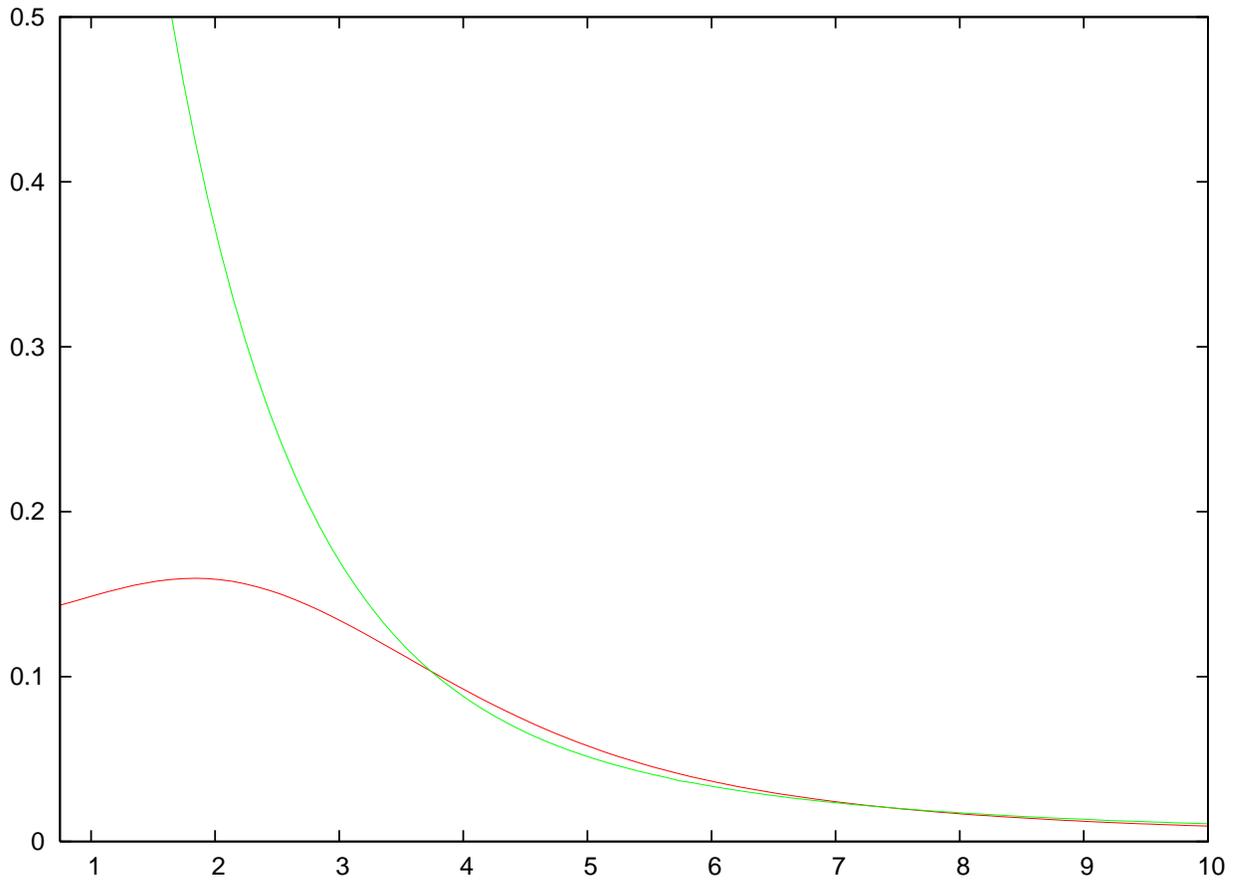}}
\caption{Unintegrated gluon distribution function defined by Balitsky-Kovchegov(red) and
McLerran-Venugopalan(green)}
\label{fig_MV_BK}
\end{figure}

It is well known that in saturation model there is two ways to define unintegrated gluon distribution
function based on $N({\bf r},Y)$: Balitsky-Kovchegov way and McLerran-Venugopalan way:
\begin{equation}
\label{BK_phi}
\phi^{BK}_A (x, q^2) =  
\frac{4 S_A C_F}{\as (2 \pi)^3} \int 
d^2 r \, e^{- i q r} \ \nabla^2_r N_G(r,x))
\end{equation}
\begin{equation}
\label{MV_phi}
\phi^{MV}_A (x, q^2) =  
\frac{4 S_A C_F}{\as (2 \pi)^3} \int 
d^2 r \, e^{- i q r} \ \frac{1}{r^2} N_G(r,x))
\end{equation}
or using rotation invariance of N(r,x) we can derive
\begin{equation}
\label{phi_cil_BK}
\phi_A (x, q^2) =  
\frac{4 S_A C_F}{\as (2 \pi)^2} \int_{0}^{\infty} 
dr \, J_0(qr) r\nabla^2_r N_G(r,x)
\end{equation}
\begin{equation}
\label{phi_cil_MV}
\phi_A (x, q^2) =  
\frac{4 S_A C_F}{\as (2 \pi)^2} \int_{0}^{\infty} 
dr \, J_0(qr) \frac{1}{r} N_G(r,x)
\end{equation}
This two definition gives similar momentum dependency for unintegrated gluon distribution function
at high momentum (i.e pure bremsstrahlung) but has a quite different behaviour at low momentum
(fig. \ref{fig_MV_BK}).

We could expect that these different definitions for gluon distribution functions with $N(r,x)$ defined throw (\ref{BK_eqn}) lead to same dependence of Cronin momentum.
Since equation (\ref{BK_eqn}) was written with fixed strong coupling in mind it is not oblivious why we even should have
different dependency for Cronin momentum from (\ref{phi_cil_BK},\ref{phi_cil_MV}).
However since initial condition already contains soft scale $\Lambda_{QCD}$ to provide right ultraviolet asymptotic for unintegrated gluon distribution function and (\ref{phi_cil_BK}),(\ref{phi_cil_MV})
have very different behaviour at infrared momenta (in case of (\ref{phi_cil_BK}) unintegrated gluon distribution function approaches to $0$ while $q \rightarrow 0$, but with (\ref{phi_cil_MV}) unintegrated gluon distribution function approaches to $\infty$)
it is clear that using (\ref{phi_cil_BK}) and (\ref{phi_cil_MV})
will lead to different behaviour for unintegrated gluon distribution function.

In saturation framework gluon production cross-section in $AB$ collisions can be written in simple $k_t$-factorized form:
\begin{equation}
\frac{d \sigma^{AB}}{d^2 q \ dy} \, = \, \frac{2 \, \as}{C_F} \,  
\frac{1}{q^2} \, \int d^2 k \, \phi_A (x_1,q^2) \, \phi_B (x_2,(q  
- k)^2)
\label{AB_cs}
\end{equation}
where $\phi_{A,B}$ is unintegrated gluon distribution of nucleus and proton and
$x_1,x_2$ are defined by equation
\begin{eqnarray}
x_1=\frac{q}{\sqrt{s}}e^{-y},\ x_2=\frac{q}{\sqrt{s}}e^{y}\,.
\end{eqnarray}
Of cause in leading log approximation (\ref{AB_cs}) reduces to the following form
\begin{equation}
\frac{d \sigma^{AB}}{d^2 q \ dy} \, = \, \frac{2 \, \as}{C_F} \,  
\frac{1}{q^2}
\left(
\phi_A(x_1,q^2)G_B (x_2,q^2)+
G_A(x_1,q^2)\phi_B (x_2,q^2)
\right)
\label{AB_cs_app}
\end{equation}
where $G(x,q^2)$ is gluon distribution function.
But it is clearly seen that (\ref{AB_cs_app}) and (\ref{AB_cs}) have very different behaviour at low momentum when
unintegrated gluon distribution have the form (\ref{phi_cil_BK}), and usage of (\ref{AB_cs_app}) should be avoided for
momentum dependency calculation.
Having at hand cross section for $A+B$ scattering Cronin ratio can be easy written
\begin{equation}
R_{AB} = \frac{\frac{d\sigma_{AB}}{dyd^2p}}{\frac{d\sigma_{pp}}{dyd^2p}}
\label{R_AB}
\end{equation}
With this Cronin momentum can be easily calculated.

\section{Numerical Methods}
There are three components we have for numerical calculation: cross section (\ref{AB_cs}), unintegrated gluon distribution
functions (\ref{phi_cil_BK}),(\ref{phi_cil_MV}), forward scattering amplitude $N(r,x)$ evolved by Balitsky-Kovchegov equation
(\ref{BK_eqn}). We address them in order.

First lets consider cross section calculation.
There is a few problems with direct numerical calculation of integral in cross-section (\ref{AB_cs}).
It should be stressed that even if unintegrated gluon distribution function defined in (\ref{phi_cil_BK})
has no singularity at $q^2=0$, (\ref{phi_cil_MV}) does.
So we should write numerical integration procedure which symmetrically avoids both singularities in (\ref{AB_cs}),
one from each gluon distribution function.
Since $\phi(k)\propto \frac{1}{k^2}$ at high momentum $k$ it is hard to calculate
integral in (\ref{AB_cs}) directly due slow downfall at high momenta.
Of cause it is not a problem in analytical calculation since this integral is a good defined one.
Since direct computation has have before-mentioned problems it is clear we should use some other method.
Moreover even if we decide to do direct computation for this integral it will cause unnecessary loss
of precision.

It is easy to calculate integral in (\ref{AB_cs}) by the following method.
Lets transform integration area from $R^2$ to something more foreseeable, like $I^2$
(numerical integration in infinite boundaries is complicated procedure by itself and should be avoided if possible).
It will be convenient if resulting integrand will have same order of magnitude in all $I^2$ and we chose transformation
which satisfy this condition.
Lets divide singularities i.e. separate integral into two pieces.
It can easily be done with following prescriptions
\begin{eqnarray}
\frac{d \sigma^{AB}}{d^2 q \ dy} = 
\frac{2 \, \as}{C_F} \frac{1}{q^2}
\int_{{\bf q}{\bf p}<q_0p} d^2 k \, \phi_A (x_1,q^2) \, \phi_B (x_2,(q  - k)^2)+
\\\nonumber
+\frac{2 \as}{C_F} \frac{1}{q^2}\
\int_{{\bf q}{\bf p}<(p-q_0)p} d^2 k \, \phi_A (x_1,q^2) \, \phi_B (x_2,(q  - k)^2)
\label{cs_decomp}
\end{eqnarray}
where $q_0$ is some value with property $0<q_0<p$ (for simplicity we'll take $q_0=\frac{p}{2}$.
Now we'll make the following parameter transformation in first term in (\ref{cs_decomp})
\begin{equation}
    (q,\psi) \rightarrow \left(\frac{{\Lambda}^2}{q^2},\frac{\psi-{\psi}_0}{\pi-{\psi}_0}\right)
\label{cs_newparm}
\end{equation}
where ${\psi}_0=arcos(min(1,(q_0/q))$
Similar transformation should be done in second therm.
It is clearly seen that equation (\ref{cs_newparm}) transform vector ${\bf q}$ into pair $(x,y)$ which belongs to $I^2$.
Of cause this transformation gives additional measure to integrand in (\ref{cs_decomp}) which equals to
$(\pi-\psi_0)\cdot \frac{2q^3}{\Lambda^2}$.
After this transformation both resulting integrals could be calculated very fast by adaptive Cubo method.

Lets consider unintegrated gluon distribution function now.
It clearly seen that $z$ integration in (\ref{phi_cil_BK}) can be easily done since $\nabla^2$ efficiently cuts off integration
area. However it is not the case for (\ref{phi_cil_MV}) since integrand falls slowly with increasing of $z$.
One of possible method for numerical calculation of (\ref{phi_cil_MV}) is just cut off $z$ integration at some big $z$ and
analytically calculate the rest. This however leads to quite complicated formula for tail term.
We choose here more simpler (from numerical point of view) method.
Since roots of Bessel function $J_0(x)$ is well known and approximately equals to
\begin{equation}
	x_i=(i+\frac{3}{4})\pi
\end{equation}
we can define set of finite integrals from $0$ to $x_{i+1}$
\begin{equation}
S_i =  
\frac{4 S_A C_F}{\as (2 \pi)^2} \int_{0}^{2x_i} 
dr \, J_0(qr) \frac{1}{r} N_G(r,x)
\end{equation}
It could be easily showed that $S_i$ have monotonic behaviour at high i.
By applying Eckler process to pair $(\frac{1}{i},S_i)$ and using polynomial extrapolation
we can calculate (\ref{phi_cil_MV}) with any accuracy.
It should be noted that using more precise value for Bessel function roots gives substantial
speedup for this procedure and therefore we'll use more precise McMahon formula which in next order gives.
\begin{equation}
	x'_i=x_i+\frac{1}{8 x_i}-\frac{31}{384 {x^3_i}};
\end{equation}

And the last and most important component is forward gluon scattering amplitude $N(r,x)$.
It should be noted what Balitsky-Kovchegov equation can be solved much simpler in momentum space
rather coordinate space due its analytical properties.
However we'll solve it in coordinate space. And we have solid reason for this.
Developed method is fast and precise enough and can be easily applied for recently proposed
enchantments of Balitsky-Kovchegov
equation where coordinate dependence of gluon forward scattering amplitude interesting by itself, and
this enchantments can not be rewritten easily in momentum space.
First we rewrite Balitsky-Kovchegov equation in terms of variable $n(r,x)=1-N(r,x)$
\begin{equation}
{dn(|{\bf r}|,y)\over dy} = -{(\as\, N_c)\over 2\pi^2}
\int d^2{\bf z}
{{\bf r}^2\over ({\bf r}-{\bf z})^2\, {\bf z}^2}
[n(|{\bf r}-{\bf z}|)-n(|{\bf r}-{\bf z}|)n(|{\bf z}|)]
\label{BK_eqn_1}
\end{equation}
It is clearly seen that method used for numerical calculation of integral in (\ref{AB_cs})
can be used here too, since integral in (\ref{BK_eqn_1}) has very similar structure,
however since $n(r,x)$ has not singularity at $r=0$ we need not exclude area around this point.
Moreover since $n(0,x)=1$ this gives some not vanishing contribution to integral
(\ref{BK_eqn_1}) we shouldn't do that.
However we could easily drop region $|z|>z_{max}$ (where $z_{max}$ some big value)  from integration
since $n(r,x)$ falls rapidly.
Therefore we choose transformation in the following form 
\begin{equation}
    (z,\psi) \rightarrow \left(\frac{a}{z-c}-b,\frac{\psi-{\psi}_0}{\pi-{\psi}_0}\right)
\label{BK_newparm}
\end{equation}
where ${\psi}_0=arcos(min(1,(r_0/z))$ and $0<r_0<r$
and constants $a,b,c$ will be chosen to follow to the following conditions
\begin{eqnarray}
\frac{-a}{c}-b=1 \\\nonumber
\frac{a}{z_{max}-c}b=0  \\\nonumber
\frac{a}{z'-c}{b}=0.5
\end{eqnarray}
where $z'$ is some characteristic scale for $n(z,y)$.
After this integration can be done by adaptive Cubo method.
Next we should decide how to do rapidity integration in (\ref{BK_eqn_1}).
It is oblivious that since $n(z,y)$ will fall with rise of $y$ rises and approach {\it asymptotically} to
zero it is not possible to use explicit integration schema here since it breaks this asymptotic behavior easily.
Therefore we'll use second order implicit integration method for integration since it gives necessary precision in
integration and does not violate asymptotic behavior.

\begin{figure}
\resizebox{1\hsize}{!}{\epsffile{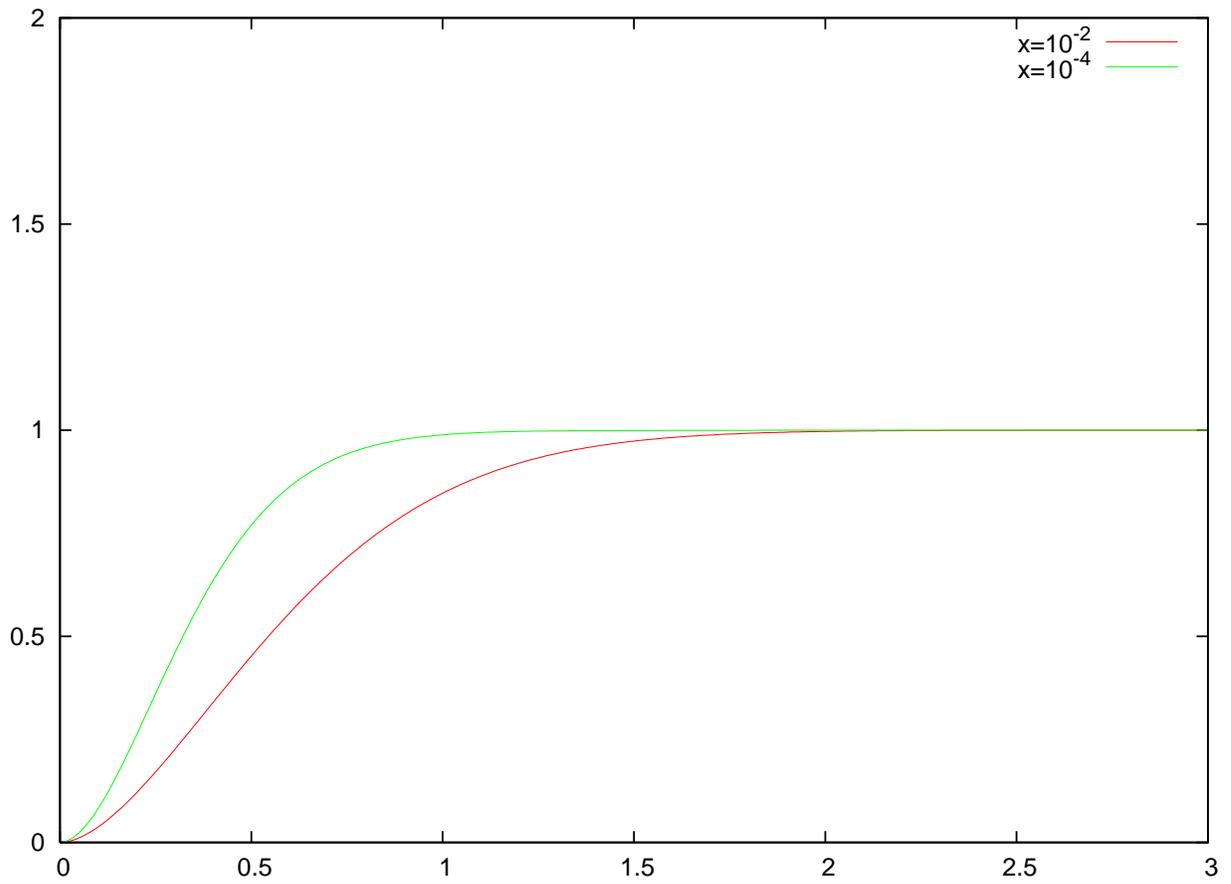}}
\caption{Forward gluon scattering amplitude at different $x$ evolved by Balitsky-Kovchegov equation}
\label{fig_NG_BK}
\end{figure}

\section{Cronin momentum in $p+A$, $d+A$ and $A+A$ collision}
There are three cases in high energy nuclear collision in which we should check Cronin momentum dependency:
$p+A$, $d+A$, $A+A$.
But first we should define few parameters and initial conditions, i.e. form of gluon distribution function for $p$, $d$
and $A$ and fix value of $\alpha_s$.
For nucleus $A$ (which we suppose to be $Au$) we'll take unintegrated gluon distribution functions in the form
(\ref{phi_cil_BK}),(\ref{phi_cil_MV}) with forward scattering amplitude evolved from initial form 
\begin{equation}
N_0(r)=1-e^{-r^2Q_s(x)^2\ln(1/r\Lambda+\epsilon)/4}
\label{NG_A_init}
\end{equation}
given at $x_0=0.01$ with initial saturation momentum $Q^2_{0s}=2Gev$ and regularization parameter $\epsilon=0.1$
(obtained results does not depend on exact value of $\epsilon$ if $0<\epsilon<1$)
For deuteron we set same initial condition as in (\ref{NG_A_init}) with saturation momentum rescaled according mass number
difference.
Finally for proton we chose simple 'bremsstrahlung' form for unintegrated gluon distribution function.
There is only $\alpha_s$ value left to set.
Since we consider Balitsky-Kovchegov equation with fixed strong coupling we can not set it using some scale.
However since (\ref{qsat_real}),(\ref{qsat_simple}) exhibit same asymptotic behaviour we can relate
value of $\lambda$ obtained in high energy region and $\bar{\alpha_s}$ with the following relation:
\begin{equation}
\lambda=\bar{\alpha}\frac{\chi(\gamma_c)}{\gamma_c}
\end{equation}
However the only known $\lambda=0.3$ was fixed by DIS data at x=0.01 \cite{DIS_fix}.
The asymptotic one should be lesser that $\lambda=0.3$ but scaling law (\ref{qsat_real}) does not contain all
terms and does not allow determine exact value of $\lambda$.
Therefore we determine $\lambda$ and consequently $\bar{\alpha}$ by the following method.
Fix some some arbitrary value $\bar{\alpha}$, calculate forward scattering amplitude evolved
by Balitsky-Kovchegov equation, determine linear component of scaling law (\ref{qsat_real}) at $x=0.01$
and rescale $\bar{\alpha}$ to fit $\lambda=0.3$.
After that we can calculate energy and rapidity dependency of Cronin momentum.

\begin{figure}
\begin{center}
\resizebox{1\hsize}{!}{\epsffile{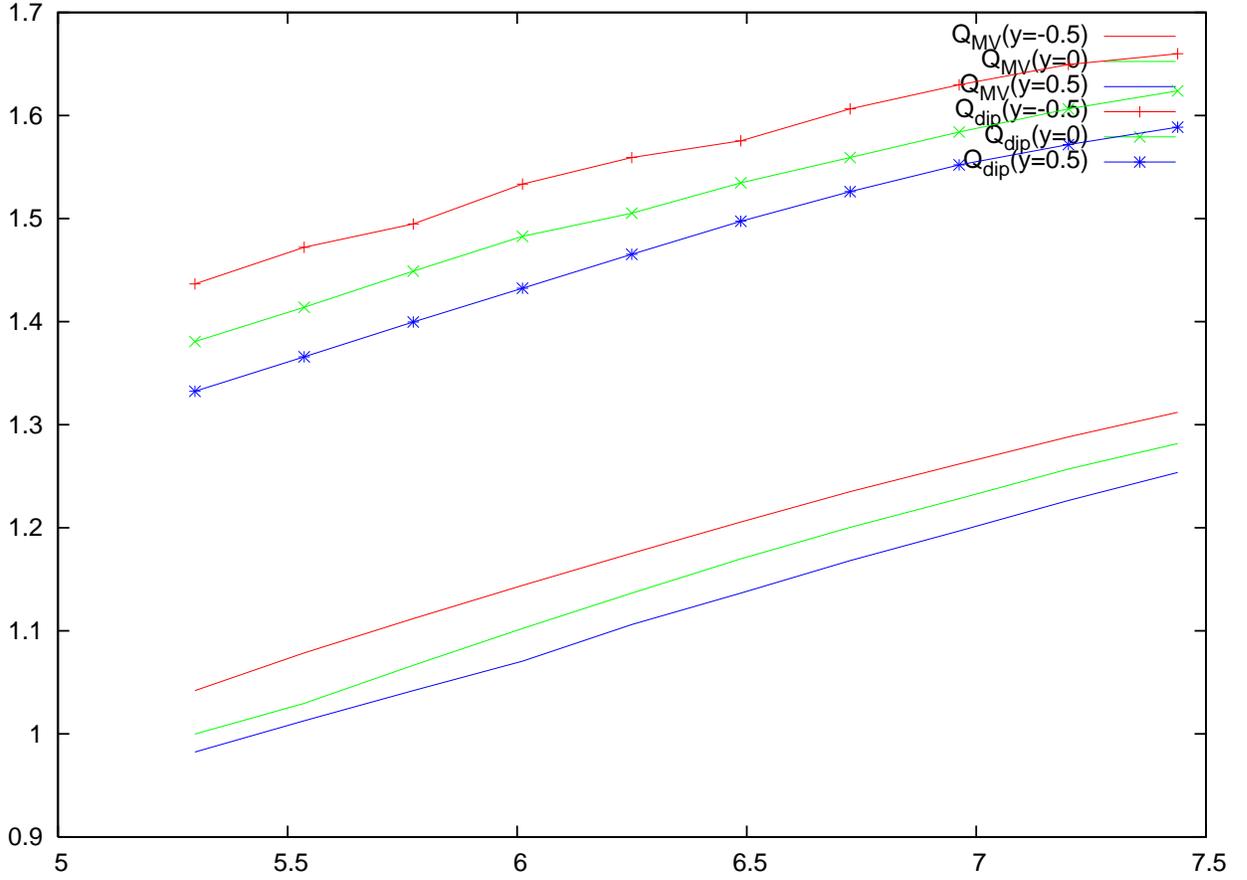}}
\end{center}
\caption{Energy logarithm dependence of Cronin momentum for $A+p$ collisions at different rapidities with cross section
calculated using gluon distribution function (\ref{phi_cil_BK}) (dotted line) and (\ref{phi_cil_MV}) (simple line)}
\label{fig_CM_s1}
\end{figure}

Energy dependency of Cronin momentum for $p+A$ collision is shown in figures
\ref{fig_CM_s1} and 4.
It is clearly seen that logarithm of Cronin momentum calculated using unintegrated gluon distribution function
(\ref{phi_cil_BK})
higher that Cronin momentum calculated using unintegrated gluon distribution function (\ref{phi_cil_MV}).
The ratio of Cronin momentum for gluon distribution functions (\ref{phi_cil_BK}), (\ref{phi_cil_MV}) is approximately
constant and equal $exp(0.37)$. The same also holds also for $d+A$ and $A+A$ collision where Cronin momentum exhibits
the same energy behaviour. In $A+p$ collision rapidity dependence of Cronin momentum mimics exactly energy dependence since gluon distribution function of proton was chosen in simple 'bremsstrahlung' form.

\begin{figure}
\begin{center}
\resizebox{1\hsize}{!}{\epsffile{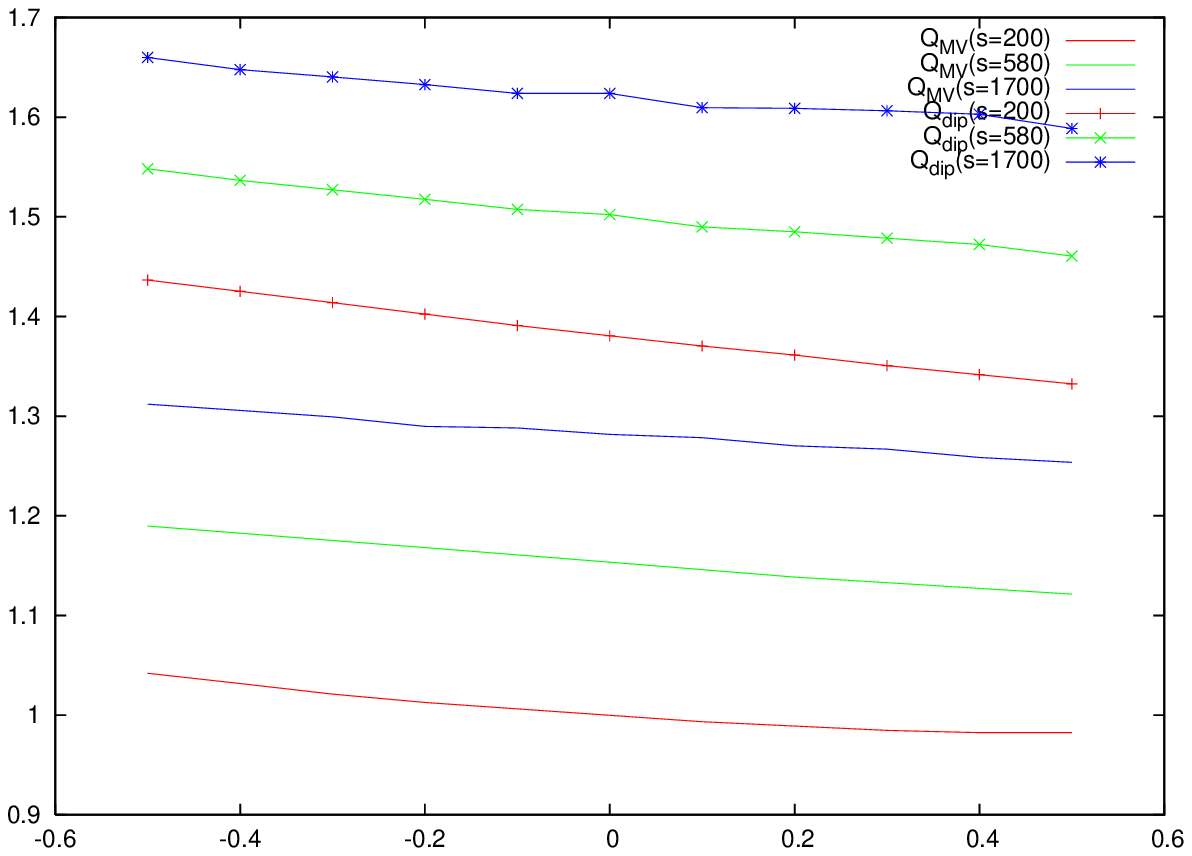}}
\end{center}
\label{fig_CM_y1}
\caption{Rapidity dependence of Cronin momentum for $A+p$ collisions at different energies with cross section
calculated using gluon distribution function (\ref{phi_cil_BK}) (dotted line) and (\ref{phi_cil_MV}) (simple line)}
\end{figure}

\begin{figure}
\begin{center}
\resizebox{1\hsize}{!}{\epsffile{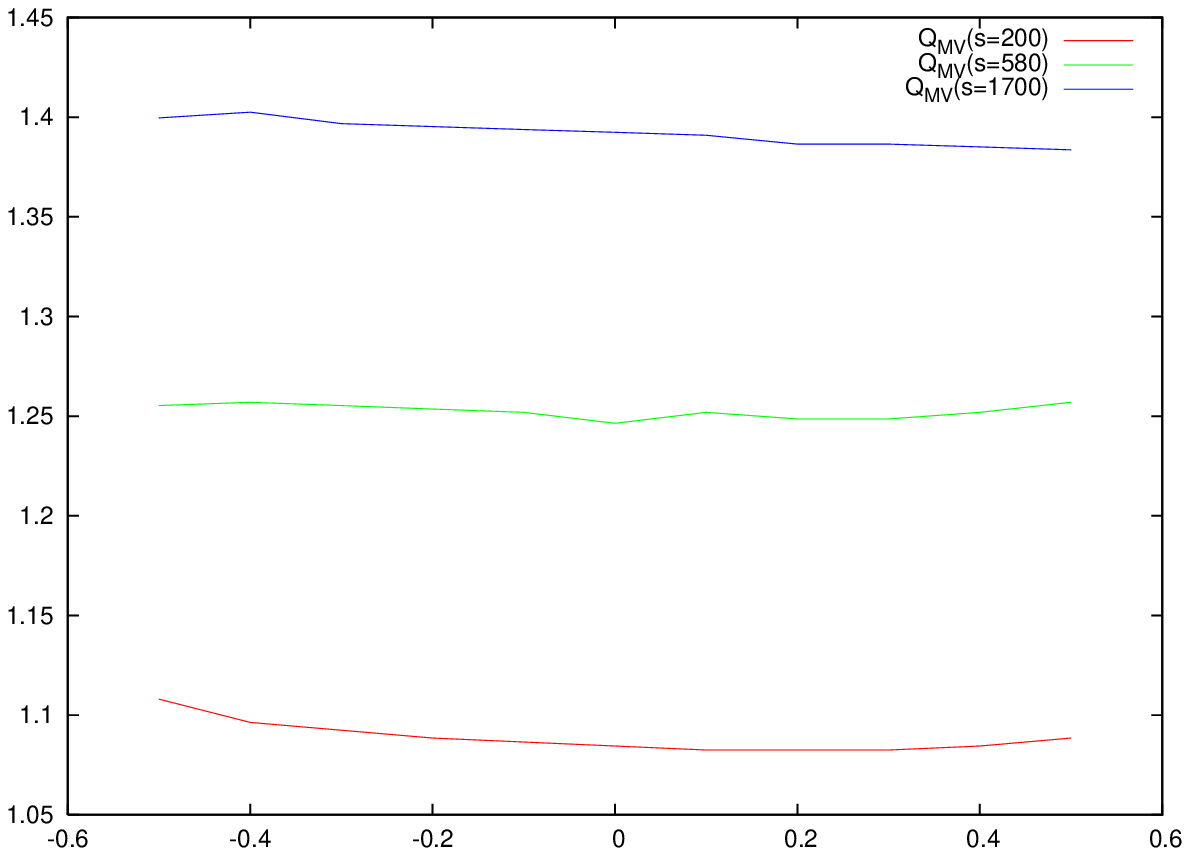}}
\end{center}
\caption{Rapidity dependence of Cronin momentum for $A+A$ collisions at different energies with cross section
calculated using gluon distribution function (\ref{phi_cil_MV})}
\label{fig_CM_y2}
\end{figure}

\begin{figure}
\begin{center}
\resizebox{1\hsize}{!}{\epsffile{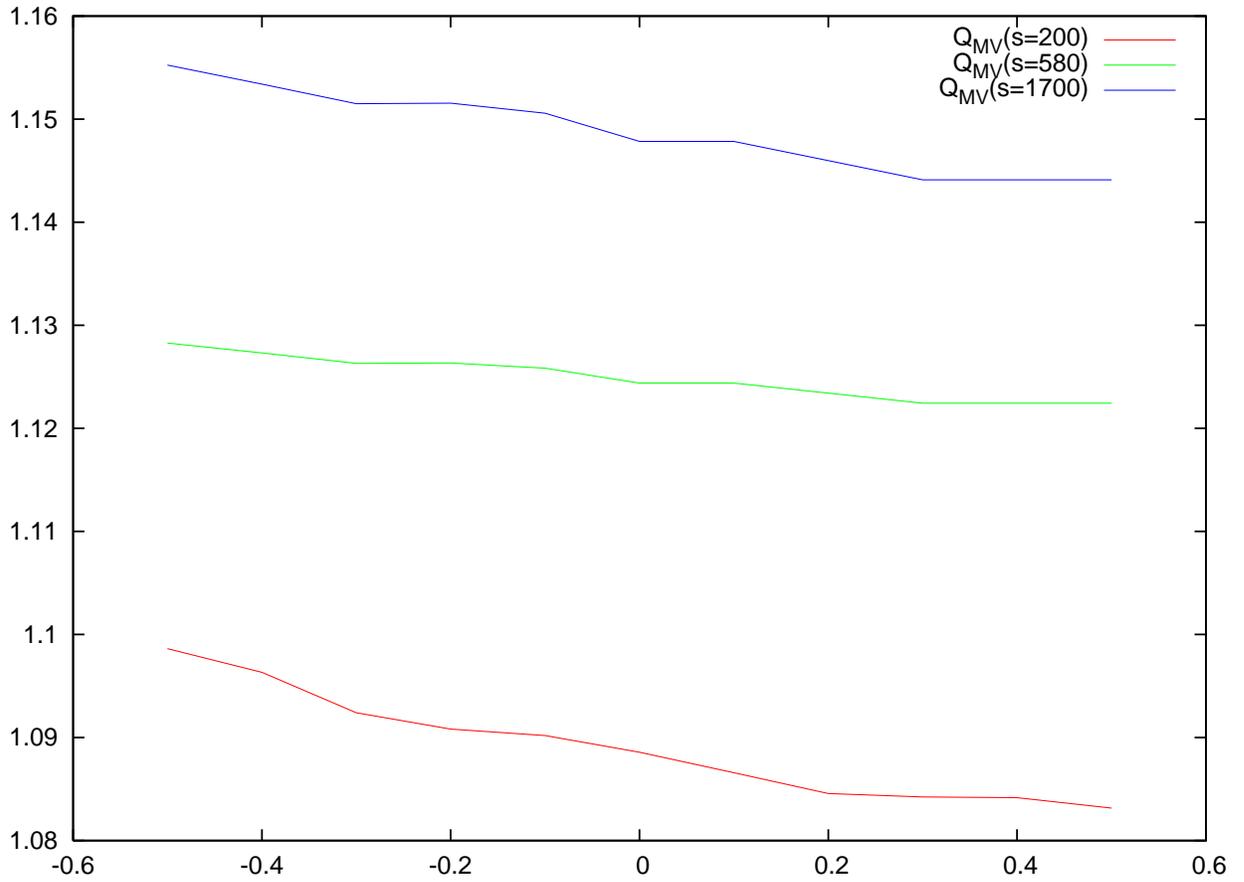}}
\caption{Rapidity dependence of Cronin momentum for $A+d$ collisions at different energies with cross section
calculated using gluon distribution function (\ref{phi_cil_MV})}
\end{center}
\label{fig_CM_y3}
\end{figure}

Rapidity dependence for $A+A$ and $A+d$ collisions is shown in figures (\ref{fig_CM_y2}) and (\ref{fig_CM_y3}).
It is clearly seen that Cronin momentum in $A+A$ collision does not have rapidity dependency which is quite
unexpected behaviour while for $A+d$ we have dependency which consistent with (\ref{CM_simple}). 

\newpage

\section{Conclusion}
Even usage of Balitsky-Kovchegov equation for forward scattering amplitude leads to
energy behavior of Cronin momentum related with scaling law by dimensional consideration.
It is clearly seen that although in simple model \cite{Abramovsky:2003xx} with scaling law (\ref{qsat_simple})
Cronin momentum calculated using different gluon distribution function have slightly different energy
behaviour \cite{Abramovsky:2003xx} in case of more natural gluon distribution function defined throw solution of
Balitsky-Kovchegov equation there is no difference between Cronin momentum energy behaviour.
It is easy to understood since Balitsky-Kovchegov equation noes not contain soft scale except in initial condition.
 
However:
\itemize{
\item{We completely ignore the fact that equation (\ref{BK_eqn}) can be written using running coupling constant.
It was shown that this modification changes greatly scaling law (\ref{qsat_real}) but does not break scaling at all.
This fact alone should set the same Cronin momentum energy behaviour for both definition of unintegrated gluon distribution.
Since this modification leads to scaling law which does not restores to (\ref{qsat_simple}) even at asymptotically
high energies it could be easily tested experimentally ether with more precise Cronin momentum rapidity dependence or with
higher energy measurement. This allows to test if a way to introduce running coupling constant
is the right one.}
\item{It was shown in \cite{BK_b_num_sol} what finite nucleus size sets yet another scale related to its radius $R$.
Since this scale is soft it should leads to different results with unintegrated gluon distribution functions (\ref{phi_cil_BK}), (\ref{phi_cil_MV}) in both fixed or running coupling case even if we consider forward scattering amplitude $N_G$
evolved by Balitsky-Kovchegov equation.}
\item{In Cronin momentum rapidity behaviour for $A+A$ collisions it is cleanly seen that there
is no rapidity dependence of Cronin momentum. However since we should take into account two different dipole
forward scattering amplitude (quark and gluon dipole) which obeys different scaling law this behaviour could be easy broken
and should be tested.}
}
\section*{Acknowledgments}
We thank A.V.Dmitriev and A.V. Popov for useful discussions. This work was supported by RFBR
Grant RFBR-03-02-16157a.

\end{document}